\documentstyle[11pt]{article}
\newtheorem{lemma}{Lemma}

\begin{document}

\title{The Garman--Klass volatility estimator revisited}

\setcounter{page}{0}

\baselineskip= 20pt

\author{
Isaac Meilijson\footnotemark \\
{\em Tel-Aviv University }}

\maketitle

\begin{abstract}
\begin{quote}

The Garman--Klass unbiased estimator of the variance per unit time
of a zero--drift Brownian Motion $B$, based on financial data that
reports for time windows of equal length the open ($OPEN$), minimum
($MIN$), maximum ($MAX$) and close ($CLOSE$) values, is quadratic in
the statistic $S_1=(CLOSE-OPEN, OPEN-MIN, MAX-OPEN)$. This
estimator, with efficiency $7.4$ with respect to the classical
estimator $(CLOSE-OPEN)^2$, is widely believed to be of minimal
variance. The current report disproves this belief by exhibiting an
unbiased estimator with slightly but strictly higher efficiency
7.7322. The essence of the improvement lies in the proposal that the
data should be compressed to the statistic $S_2$ defined on $W(t)=
B(0)+[B(t)-B(0)] \mbox{sign}[(B(1)-B(0)]$ as $S_1$ was defined on
the Brownian path $B(t)$. The best $S_2$--based quadratic unbiased
estimator is presented explicitly. The Cram\'{e}r--Rao upper bound
for the efficiency of unbiased estimators is $8.471$. It corresponds
to the large-sample efficiency of Maximum Likelihood estimators.
This bound cannot be attained because the distribution is not of
exponential type.

Regression-fitted quadratic functions of $S_2$ (with mean $1$)
markedly out-perform those of $S_1$ when applied to random walks
with heavy-tail-distributed increments. Performance is empirically
studied in terms of the tail parameter.

\medskip

{\bf Keywords and phrases:} Garman--Klass, Brownian Motion,
volatility, estimation.

{\bf MSC2000:} 62F10, 62P05
\end{quote}
\end{abstract}

\footnotetext{Research conducted on a sabbatical visit to Columbia
University, $2008$}

\section{Introduction}

As stressed repeatedly (see Magdon-Ismail \& Atiya
(2001)), volatility estimators of financial data ought to have as
small a variance as possible, because volatilities change over time,
so past data have decaying importance. The celebrated Garman--Klass
(1980) variance estimator, introduced almost three decades ago,
achieves better accuracy in estimating $\sigma^2$ than the
classical, natural estimator {\em average} $(CLOSE-OPEN)^2$ does in
seven times the observation period. This unbiased variance estimator
is the minimum-variance unbiased quadratic function of the spreads
$c=CLOSE-OPEN, h=MAX-OPEN, l=MIN-OPEN$ (for {\em close, high, low}).
These data $S_1=(c, h, l)$ can be compressed without loss of
sufficiency.

\noindent {\bf A coarser (but incomplete) sufficient statistic}.
Consider the triple $S_2=(C, H, L)$ where $C = |c| \ , (H,L) = (h,l)$ if
$c>0$, while $(H,L) = -(l,h)$ if $c<0$. Without
loss of relevant information about the variance,
the Brownian Motion trajectory $\{B(t) \ ; \ t \in (0,1)\}$ may be replaced by
the flipped path $\{W(t) \ ; \ t \in (0,1)\}$, defined as
$W(t)=B(0)+[B(t)-B(0)] \mbox{sign}(B(1)-B(0))$. That is, the
three interval lengths $(-L,C,H-C)$, in fact
the further compression
$(C,\min(-L,H-C),\max(-L,H-C))$,
determined by $(c,h,l)$, carry all relevant information contained in
$(c,h,l)$ about $\sigma^2$, but {\em do not determine} $(c,h,l)$. Although
intuitively clear after some thought, sufficiency
of $(C,\min(-L,H-C),\max(-L,H-C))$ can be formally inferred from Siegmund's
(1985) representation displayed as (\ref{Siegmformula}) in the
sequel. The Rao--Blackwell theorem
(Blackwell (1947), Rao (1946)) claims that under these conditions,
for every $S_1$-based unbiased estimator of some arbitrary parameter
there is an $S_2$-based unbiased estimator with smaller variance --
strictly smaller unless the two coincide. As will be seen, the
Garman--Klass estimator is a function of $S_2$, so the Rao-Blackwell
improvement leaves it invariant. However, the Garman--Klass
estimator, best among the quadratic function of $S_1$, is not best
possible as a function of $S_2$. Had $S_2$ been a complete minimal
sufficient statistic, Garman--Klass and the proposed estimator would
have equally been the UMVUE (uniformly minimum variance unbiased
estimator) of the parameter. However, $C^2$ and $2[(H-C)^2+L^2]$ are
different unbiased estimators of $\sigma^2$. Hence, $S_2$ (whether
minimal sufficient or not) is not complete. Loose some, win some: we
will only conjecture rather than claim optimality of the proposed
$S_2$--based quadratic unbiased estimator of $\sigma^2$; on the
other hand, the exchangeability property under which $(-L,C,H-C)$
and $(H-C,C,-L)$ are identically distributed, justifies searching
for the best quadratic function of $(-L,C,H-C)$ among those that are
linear combinations of four rather than six quadratic terms.

\noindent {\bf Four basic quadratic unbiased variance estimators}.
\begin{equation} \label{fourterms}
\hat{\sigma}_1^2 = 2 [(H-C)^2+L^2] \ , \ \hat{\sigma}_2^2 = C^2 \ ,
\ \hat{\sigma}_3^2 = 2(H-C-L)C \ , \ \hat{\sigma}_4^2 = -{{(H-C)L}
\over {2 \log(2)-{5 \over 4}}}
\end{equation}
The rationale for the somewhat bizarre coefficients is that each of
these four terms is an unbiased estimator of $\sigma^2$, with
respective variances
\begin{equation} \label{fourtermsvariance}
\mbox{Var}(\hat{\sigma}_1^2) = 0.797943 \sigma^4 \ , \
\mbox{Var}(\hat{\sigma}_2^2) = 2 \sigma^4 , \
\mbox{Var}(\hat{\sigma}_3^2) = 0.504753 \sigma^4 \ , \
\mbox{Var}(\hat{\sigma}_4^2) = 1.004876 \sigma^4
\end{equation}


\noindent {\bf The proposed variance estimator vis \`{a} vis
Garman--Klass}. The proposed estimator $\hat{\sigma}^2=\sum_1^4
\alpha_i \hat{\sigma}_i^2$ assigns to these four terms respective
weights
\begin{equation} \label{thealphas}
\alpha_1=0.273520 \ , \ \alpha_2=0.160358 \ , \ \alpha_3=0.365212 \
, \ \alpha_4=0.200910
\end{equation}
and achieves variance $\mbox{Var}(\hat{\sigma}^2)=0.258658 \sigma^4$.
The Garman--Klass estimator \cite{GarKlass}
\begin{equation} \label{GarKlassest}
\hat{\sigma}_{GK}^2=0.511 (h - l)^2 - 0.019(c(h + l)-2 h l) - 0.383
c^2
\end{equation}
happens to pool these four basic estimators too, so the Rao--Blackwell
theorem does not rule out the possibility that it
coincides with $\hat{\sigma}^2$. However, as argued earlier, the two do not agree, and
$\hat{\sigma}_{GK}^2=\sum_1^4 \beta_i \hat{\sigma}_i^2$ pays a price
for being quadratic in $(c,h,l)$. Its coefficients are given by
\begin{eqnarray}
\beta_1 &=& {0.511 \over 2} = 0.2555 \nonumber \\
\beta_2 & = & 0.511-0.383-0.019=0.1090 \nonumber \\
\beta_3 & = & 0.511-{0.019 \over 2}=0.5015 \nonumber \\
\beta_4 & = & 2(0.511 - 0.019) (2 \log(2) -{5 \over 4})=0.1340
\label{GarKlass}
\end{eqnarray}
\noindent that achieve $\mbox{Var}(\hat{\sigma}_{GK}^2)=0.27
\sigma^4$.

\noindent {\bf Maximum Likelihood variance estimators and Fisher
information}. In principle, giving up on the requirement of
unbiasedness, the computer--intensive maximum likelihood estimator
(MLE) of $\sigma^2$ by Magdon-Ismail \& Atiya
(2001) could have been a competitor, since MLE's are functions of
any sufficient statistic. However, this estimator is based on
$(h,l)$ rather than on $(c,h,l)$. Magdon-Ismail \& Atiya report that
their estimator has variance slightly higher than Garman--Klass'.

The joint generating function of $(c,h,l)$ is presented by Garman \&
Klass as an infinite series, from which these authors derived all
pertinent second and fourth degree moments.

Ball \& Torous (1984)
developed an infinite--series formula for the joint density of
$(c,h,l)$ and used it to construct numerically the MLE of
$\sigma^2$. They report estimated efficiency of the MLE for a
selection of sample sizes, basing each value on a simulation sample
size of $1000$ runs, a great achievement in $1984$, but insufficient
for delicate comparisons. An attempt at numerical evaluation of the
Fisher information, based on the Ball \& Torous expression for the
joint density, disclosed that their formula seems to have a
missprint. This joint density was re-derived based on the formula by
Siegmund quoted earlier, exhibited as (\ref{Siegmformula}) in the
sequel. The inverse of the Fisher information is the Cram\'{e}r--Rao
lower bound for the variance per time--window of any unbiased
estimator of $\sigma^2$, for any sample size. It is also the
asymptotic variance of the (not necessarily unbiased) MLE of
$\sigma^2$. Its value turns out to be $0.2361$. This is the
benchmark with which Garman--Klass' $0.27$ and the proposed
estimate's $0.258658$ variances should be compared.

\noindent {\bf The Cram\'{e}r--Rao bound $0.2361$ is not attained by
unbiased variance estimators: disproving exponentiality of a family
of distributions}. Under proper regularity assumptions (see Joshi
(1976)
), the Cram\'{e}r--Rao bound is attained if and
only if there is a linear relationship between the estimator and the
score function (derivative with respect to the parameter of the
logarithm of the density). However, for this to happen, there must
exist a linear relationship between the score functions evaluated at
different values of the parameter. It was ascertained numerically
that this is not the case. In other words, the model is not of
exponential type. We don't know whether the sufficient statistic
$S_2$, shown above not to be complete, is minimal sufficient. As a
result of all of these considerations, the proposed estimator may
not be of minimal variance.

Since both the proposed and Garman--Klass' estimators are averages
over time--windows, their variances per time--window are independent
of sample size. It is conceivable, and Ball \& Torous have provided
evidence in this direction, that the MLE has variance per
time--window that decreases as the sample size increases, so for
small sample sizes the proposed estimator has in practice no
competitor.

Moreover, since the BM model doesn't really hold in practice, a
broader contribution of this paper is the introduction of more
efficient quadratic statistics on which to base practical
estimators. Simulation results for random walks with $t$-distributed
increments are reported in Section \ref{simulation}.

\section{Derivation}

Following the steps of Garman \& Klass, all second and fourth order
moments of $(C,L,H)$ will be identified. Some of these will be
quoted from Garman \& Klass, some will be derived once the joint
densities of $(C,H)$ and $(C,L)$ are explicitly presented, and some
will require some additional argument. Although it would perhaps be
more natural to work only with the exchangeable variables
$\Delta=H-C$ and $\delta=-L$, work will be performed on the
variables $H$ and $L$ as well, in order to link more easily with
Garman \& Klass' triple $(c,h,l)$.

\subsection{The joint densities of $C$ and each of $H$ and $L$: four
unbiased estimators}

Assume throughout the computations that the drift is $0$ and the
variance per unit time is $1$. Thus, $E[C^2]=E[c^2]=1$.

By a common reflection argument, $BM$ reaching at least as high as
$x>0$ and ending up at $y = x-(x-y) \in (0,x)$ is tantamount to
ending up at $x+(x-y)$. Or, \linebreak $P(H>x,C \in [y,y+dy])=P(C
\in [2x-y,2x-y+dy])=2 \phi(2x-y)dy$, where $\phi(\cdot)={1 \over
\sqrt{2 \pi}} \exp\{-{1 \over 2}(\cdot)^2\}$ is the standard normal
density function (see Siegmund or expression (\ref{Siegmformula}) in
the Appendix for a generalization to $(C,H,L)$).

Similarly, $P(L<z,C \in [y,y+dy])=P(C \in
[2z-y,2z-y+dy])=2\phi(2z-y)$. Hence, the joint density of $H$ and
$C$ is
\begin{equation} \label{densHC}
f_{H,C}(x,y)=4 (2x-y) \phi(2x-y) \ , \ 0 < y < x
\end{equation}
and that of $L$ and $C$ is
\begin{equation} \label{densLC}
f_{L,C}(z,y)=4 (y-2z) \phi(y-2z) \ , \ z < 0 < y
\end{equation}

These joint densities, essentially re-phrasings of a well known
formula for the joint density of $(h,h-c)$ (see Yor
(1997)), lead to the first four of the following five second
moments. The fifth is taken from Garman \& Klass.
Details are omitted. $E[C^2]$=1 by assumption.
\begin{equation} \label{2moments}
E[H^2]={7 \over 4} \ , \ E[L^2]={1 \over 4} \ , \ E[C H]={5 \over 4}
\ , \ E[C L]=-{1 \over 4} \ , \ E[HL]=1-2 \log(2)
\end{equation}

As a corollary,
\begin{lemma} \label{allunbiased}
The variance estimators $\hat{\sigma}_i \ , \ i=1,2,3,4$ {\em (}see
{\em (}\ref{fourterms}{\em ))} are unbiased.
\end{lemma}

Seshadri's (1988)
theorem that $2 h (h-c)$ is
exponentially distributed with mean $1$, and is independent of $c$,
implies that $2 H (H-C)$ is exponentially distributed with mean $1$,
and is independent of $C$. This is so, simply because the
conditional distribution of $(h,c)$ given that $c>0$ is the
(unconditional) distribution of $(H,C)$.

Of course, the same applies to $2 l (l-c)$ and $2 L (L-C)$. However,
$2 H (H-C)$ and $2 L (L-C)$ are dependent (identities
(\ref{4momentsGK}) yield correlation $1+{7 \over 2} \zeta(3)-8
\log(2)=-0.3380$ between the two), and dependent given $C$.

Otherwise, it would have been very easy to sample $(C,H,L)$ triples.
As things stand, it is easy to sample pairs $(c,h)$ (and $(c,l)$) or
$(C,H)$ (and $(C,L)$), by independently sampling $c$ and $h (h-c)$.
A practical approximate method to sample $(C,H,L)$ triples is to
sample $(C',H')$ correctly, then make the wrong choice $L'=C'-H'$,
not on $[0,1]$ but on each of the $N$ sub-intervals $[{{i-1} \over
N},{i \over N}]$. The construction is correct except if $H$ and $L$
are attained in the same sub-interval, the probability of which
decreases fast as $N$ increases. Instead of letting $L'=C'-H'$,
other copulas may be used, to better approximate features of the
joint distribution of $(C',H',L')$.

\subsection{The $MLE$'s of $\sigma^2$ based on $(C,H)$ and on $(C,L)$ are unbiased}

It may be of interest to notice that (\ref{densHC}) (resp.
(\ref{densLC})), reinterpreted as  $f_{H,C}(x,y;\sigma)=4 {{2x-y}
\over \sigma^3} \phi({{2x-y} \over \sigma})$, identifies the $MLE$
of $\sigma^2$ based on $(C,H)$ (resp. $(C,L)$) as the average over
the sample of ${1 \over 3} (2 H-C)^2= {1 \over 3} C^2+{1 \over 3}[4
(H-C)^2]+{1 \over 3}[4 C (H-C)]$ and ${1 \over 3} (2 L-C)^2= {1
\over 3} C^2+{1 \over 3}[4 L^2]+{1 \over 3}[-4 C L]$. The average of
the two, the simple average of the first three unbiased estimators
in (\ref{fourterms}), achieves variance $0.3694$, above
Garman--Klass'.

\subsection{The fourth moments of $(C,H,L)$}

The following fourth moments are derived from the joint densities of
$(H,C)$ and $(L,C)$. $E[C^4]=3$ is Gaussian kurtosis.
\begin{eqnarray}
E[H^4]&=&{93 \over 16} \ , \ E[L^4]={3 \over 16} \ , \ E[C H^3]={147
\over 32} \ , \ E[C L^3]=-{3 \over 32} \nonumber \\
E[C^3 H]&=&{27 \over 8} \ , \ E[C^3 L]=-{3 \over 8} \ , \ E[C^2 H^2
]={31 \over 8} \ , \ E[C^2 L^2]={1 \over 8} \label{4moments}
\end{eqnarray}

The following fourth moment information is taken from Garman \&
Klass. $\zeta$ is Riemann's zeta function, with
$\zeta(3)=\sum_{k=1}^\infty {1 \over k^3} \approx 1.2020569$.
\begin{eqnarray}
E[H^2 L^2]&=&E[h^2 l^2]=3-4 \log(2) \nonumber \\
E[C^2 H L]&=&E[c^2 h l]=2-2 \log(2)-{7 \over 8}
\zeta(3) \nonumber \\
E[H^3 L]+E[H L^3] &=& E[h l (h^2+l^2)]=6-6 \log(2) -{9 \over 4} \zeta(3) \nonumber \\
E[C H^2 L]+E[C H L^2]&=&E[c h l (h+l)]={9 \over 2} -4 \log(2)-{7
\over 4} \zeta(3) \label{4momentsGK}
\end{eqnarray}

There is one more $(C,H,L)$-based fourth moment needed, whose value
does not follow from Garman \& Klass'.
\begin{lemma} \label{HL2C}
$E[C H L^2] =\zeta(3)/16-2 \log(2)+{47 \over 32} \approx 0.1575842$.
\end{lemma}

A proof of Lemma \ref{HL2C} can be found in the Appendix. Large sample
empirical estimation of $E[C H L^2]$ gave $0.15762$, yielding
$\mbox{Var}(\hat{\sigma}_4^2)$ very close to $1$. Had $E[C H L^2]$ been
equal to $\log(2)(3 - 4 \log(2)) \approx 0.15763$ (initial conjecture),
$\mbox{Var}(\hat{\sigma}_4^2)$ would have been exactly $1$.

From all the fourth moments above,
\begin{eqnarray}
E[C^4]&=&3 \nonumber \\
E[\delta^4]&=&E[L^4]={3 \over 16} \nonumber \\
E[C \delta^3]&=&-E[C L^3]={3 \over 32} \nonumber \\
E[C^2 \delta^2]&=&E[C^2 L^2]={1 \over 8} \nonumber \\
E[C^3 \delta]&=&-E[C^3 L]={3 \over 8} \nonumber \\
E[C^2 \Delta \delta]&=&E[C^3 L]-E[C^2 H L]=2 \log(2)+{7 \over 8}
\zeta(3)-{19 \over 8} \nonumber \\
E[C \Delta \delta^2]&=&E[C H L^2]-E[C^2 L^2]=E[C H L^2]-{1 \over 8}
\nonumber \\
& = & \zeta(3)/16-2 \log(2)+{43 \over 32}
\nonumber \\
E[\Delta^2 \delta^2]&=&E[H^2 L^2]+E[C^2 L^2]-2 E[C H L^2] \nonumber
\\
& = & {3 \over 16}-{\zeta(3) \over 8}
\nonumber \\
2 E[\Delta^3 \delta] &=& E[\Delta^3 \delta]+E[\Delta
\delta^3]= -(E[H^3 L]+E[H L^3]) \nonumber \\
&+&E[C^3 L]+E[C L^3]-3 E[C^2 H L]+3 E[C H^2 L]
\nonumber \\
&=& 6 \log(2)-{9 \over 16} \zeta(3) - {27 \over 8} \label{4moments}
\end{eqnarray}

\subsection{The covariance matrix of the four basic estimators}

Let $\Sigma$ stand for the covariance matrix of the four basic
estimators. Their variances are on the diagonal, their covariances
off the diagonal.

Applying the formulas of the previous sub--section, the variances of
the basic estimators $\hat{\sigma}_i^2$ (see (\ref{fourterms})) are
\begin{eqnarray}
\Sigma(1,1)=\mbox{Var}(\hat{\sigma}_1^2)&=&8 (E[\delta^4]+
E[\Delta^2 \delta^2])-1
= 2-\zeta(3) =
0.797943
\nonumber \\
\Sigma(2,2)=\mbox{Var}(\hat{\sigma}_2^2) &=& 3-1=2 \nonumber \\
\Sigma(3,3)=\mbox{Var}(\hat{\sigma}_3^2) &=& 8 (E[C^2 \delta^2]+
E[C^2 \Delta
\delta])-1 = 8 [ \log(4)+{7 \over 8} \zeta(3)-{9 \over 4}] -1 \nonumber \\
& = & 0.504753 \nonumber \\
\Sigma(4,4)=\mbox{Var}(\hat{\sigma}_4^2) &=& {E[\Delta^2 \delta^2]
\over {({\log(4)-{5 \over 4}})^2}}-1
= {{{3 \over 16} -{\zeta(3) \over 8}} \over
{({\log(4)-{5 \over 4}})^2}}-1 = 1.004876 \label{variances4}
\end{eqnarray}
The covariances of the basic estimators are
\begin{eqnarray}
\Sigma(1,2)=\mbox{Cov}(\hat{\sigma}_1^2,\hat{\sigma}_2^2)&=& 4 E[C^2 \delta^2]-1
=  -{1 \over 2} \nonumber \\
\Sigma(1,3)=\mbox{Cov}(\hat{\sigma}_1^2,\hat{\sigma}_3^2)&=& 8 E[C \delta^3] +
8 E[C \Delta \delta^2]-1  = {{21+ \zeta(3)} \over 2}-16 \log(2) \nonumber \\
& = &
0.010674 \nonumber \\
\Sigma(1,4)=\mbox{Cov}(\hat{\sigma}_1^2,\hat{\sigma}_4^2)&=&{{4
E[\Delta \delta^3]}\over {\log(4)-{5 \over 4}}} -1 ={{12 \log(2)-{27 \over
4}-{9 \over 8} \zeta(3)}
\over {\log(4)-{5 \over 4}}} -1 \nonumber \\
& = & .580786 \nonumber \\
\Sigma(2,3)=\mbox{Cov}(\hat{\sigma}_2^2,\hat{\sigma}_3^2)&=&4 E[C^3 \delta]-1  =
{1 \over 2} \nonumber \\
\Sigma(2,4)=\mbox{Cov}(\hat{\sigma}_2^2,\hat{\sigma}_4^2)&=&{{E[C^2
\Delta \delta]} \over {\log(4)-{5 \over 4}}} -1 ={{{7 \over 8}
\zeta-{9 \over 8}} \over
{\log(4)-{5 \over 4}}}  =-.537074 \nonumber \\
\Sigma(3,4)=\mbox{Cov}(\hat{\sigma}_3^2,\hat{\sigma}_4^2) & = & {{4
E[C \Delta^2 \delta]} \over {\log(4)-{5 \over 4}}} -1 =
{{{\zeta(3) \over 4}+{43 \over 8}-8 \log(2)} \over {\log(4)-{5 \over 4}}} - 1
\nonumber \\
& = & -.043711 \label{covariances}
\end{eqnarray}

\subsection{Derivation of the proposed estimator}

Letting $\alpha$ (see (\ref{thealphas})) stand for the weights assigned to the basic
estimators, the weighted sum has variance $\alpha^T \Sigma \alpha$
and mean $\alpha^T {\bf 1}$. Using a Lagrange multiplier to constrain
the mean to be $1$, minimal variance is achieved at $\alpha =
{{\Sigma^{-1} {\bf 1}} \over {{\bf 1}^T \Sigma^{-1} {\bf 1}}}$,
yielding the weights displayed in (\ref{thealphas}). The variance of
the proposed estimator is ${1 \over {{\bf 1}^T \Sigma^{-1} {\bf
1}}}=0.258658$, with corresponding efficiency ${2 {{\bf 1}^T \Sigma^{-1} {\bf
1}}}=7.73221$.

\section{Heavy tailed random walks - simulation results}
\label{simulation}

As is commonly observed in financial data, the logarithmic
increments of returns have power-law tails, at least in the visible
range, with tail parameter around $3$. This means finite variance
but infinite variance of the usual empirical variance estimators.
Suppose that the basic process on which (Open, Close, Min, Max) data
is reported per time window is a random walk with $t$-distributed
increments. A simulation analysis will now be reported, in which the
number of increments of the random walk per time window is $10, 30$
and $50$, and the degrees of freedom ($df$) range from $1.5$ to $5$
with step size $0.5$. Minimum sum-of-squares quadratic functions
with mean $1$ of the $S_1$ and $S_2$ statistics were fitted by
Regression, with sample size $10^5$: the regression coefficients
were identically calibrated so that the predictor of unity has mean
$1$ in each such sample. Each such Regression was repeated $100$
times, and the averages of the corresponding regression coefficients
and overall ''variances" were recorded. Of course, second moments
are finite only for $df>2$ and fourth moments are finite only for
$df>4$, but the empirical study seems instructive. A sample of size
$10^5$ from the sum of $N=50$ $t_{\{df=3\}}$-distributed random
variables typically displays lighter tails than $df=3$ would entail.
Table 1 reports the empirical minimum variance of the quadratic
functions, and Table 2 reports the coefficients of the building
blocks of expression (\ref{fourterms}) that yield the
minimum-variance quadratic function for each case. These building
blocks have expectation $1$ for Brownian Motion but not for random
walk, so their coefficients need not add up to unity. Table 1
displays performances similar to those derived for Brownian Motion
for moderate $df$, fast deteriorating when $df$ decreases, in which
case $S_2$ data progressively outperforms $S_1$ data. $S_2$ data
yields lower variances than $S_1$ data throughout the range, as well
as for uniform and double exponentially distributed increments,
although the difference in variance in these light-tail cases is as
small as for BM.

\begin{center}
{\bf Table 1. Minimum variance of mean-$1$ quadratic functions of
$S_1$ and $S_2$ data}
\end{center}
\begin{center}
\begin{tabular}{c|c|c|c|c|c|c} \hline
df \ N & $10$, $S_2$ & $10$, $S_1$ & $30$, $S_2$ & $30$, $S_1$ &
$50$, $S_2$ & $50$, $S_1$ \\ \hline
    1.5 &  16.2403 &  51.0366 &   8.3438 &  32.4697 &   6.5322
 &   28.3950 \\
    2.0 &   4.8444 &   6.6039 &   2.6532  &  3.8327 &  2.1972
 &   3.2252 \\
    2.5 &   2.5864 &   2.8365 &   1.4297  & 1.5529  & 1.1718
   & 1.2627 \\
    3.0 &   1.7359 &   1.8038 &   0.9527  & 0.9782  & 0.7630
 &   0.7788 \\
    3.5 &   1.2334 &   1.2746 &   0.6809  & 0.6991  & 0.5467
   & 0.5624 \\
    4.0 &   0.9469 &   0.9776 &   0.5409  & 0.5585  & 0.4532
 &   0.4686 \\
    4.5 &   0.7864 &   0.8124 &   0.4792  & 0.4957  & 0.4094
   & 0.4239 \\
    5.0 &   0.7071 &   0.7296 &   0.4473  & 0.4629  & 0.3896
 &   0.4037 \\
 $\infty$ & 0.4679 & 0.4826 & 0.3630 & 0.3765 & 0.3369 & 0.3496 \\
 \hline
 $\infty, N=\infty$ & & & & & 0.2587 & 0.27 \\
\end{tabular}
\end{center}

It is of interest to observe how does $S_2$ outperform $S_1$ data
for low $df$. Table 2 shows that the role of $C$ is downplayed or
even dampened in favor of those of $H-C$ and $-L$, gradually
incorporating $C$ into the Regression as $df$ increases. The
rationale for this is that the tail parameter of sums of i.i.d. data
is the same as that of the summands, whereas the tail parameter of
extrema is the sum of those of the summands. This makes $C$
theoretically as heavy tailed as each increment, but makes $H-C$ and
$-L$ have lighter tails than the increments. In contrast, the $[h,
c, l]$ data of statistic $S_1$ is less able to split variables into
light tail and heavy tail components. Although $h-|c|-l=H-C-L$, the
insistence on resorting to quadratic functions leaves it out of the
$S_1$ game. Still, both statistics seem to work fairly well even
under low $df$. In contrast to the variances $2.1972$ or $3.2252$
for $df=2$, $0.7630$ or $0.7788$ for $df=3$ and $0.4532$ or $0.4686$
for $df=4$ (see $N=50$ in Table 1), the calibrated $C^2$ has
respective empirical variance above $5000$, $16$ and $2.5$,
converging reasonably fast ($2+{6 \over {(df-4) N}}$) to $2$
thereafter.

\begin{center}
{\bf Table 2. Coefficients of the minimum variance mean-$1$
quadratic function of $S_2$ data for $N=50$ increments per time
window}
\end{center}
\begin{center}
\begin{tabular}{c|c|c|c|c} \hline
df & $2((H-C)^2+L^2)$ & $C^2$ & $2(H-C-L)C$ & ${-(H-C)L \over 2\log(2)-{5/4}}$ \\
\hline
    1.5 &   0.0209 &  -0.0000 &  0.0010  &   0.1724   \\
    2.0 &   0.1358 &  -0.0004 &   0.0352 &   0.1561 \\
    2.5 &   0.1745 &  -0.0034 &   0.1573 &   0.1215 \\
    3.0 &   0.1827 &   0.0140 &   0.2461 &   0.1149 \\
    3.5 &   0.2006 &   0.0666 &   0.2460 &   0.1228 \\
    4.0 &   0.2185 &   0.1081 &   0.2442 &   0.1317 \\
    4.5 &   0.2335 &   0.1271 &   0.2620 &   0.1399 \\
    5.0 &   0.2480 &   0.1395 &   0.2781 &   0.1473 \\
$\infty$ & 0.3974 & 0.2321 & 0.4390 & 0.2245 \\ \hline
$\infty, N=\infty$ & 0.2736 & 0.1604 & 0.3652 & 0.2009 \\
\end{tabular}
\end{center}

\section{Appendix - proof of Lemma \ref{HL2C}}

For the sake of conciseness, the tedious integration to be presented
will be restricted to the identification of $E[C H L^2]$, although,
in principle, more general joint moments and moment generating
function of $(C,H,L)$ could have been identified.

Consider the infinitesimal event $\{BM(1) \in (\xi,\xi+d\xi) \ , \
BM(s) \in (a,b) \ , \ \forall s \in [0,1] \}$, where $a <
\min(\xi,0) \le 0 \le \max(\xi,0) < b$. By Siegmund's Corollary
3.43, its probability $Q(\xi,a,b) d\xi$ is as follows
\begin{equation} \label{Siegmformula}
Q(\xi,a,b) = \sum_{j=-\infty}^{\infty} \{\phi(\xi-2 j
(b-a))-\phi(\xi-2 a - 2 j (b-a))\}
\end{equation}

The joint density $f_{c,h,l}(\xi,a,b)$ is (minus) the mixed second
derivative of $Q$ with respect to $a$ and $b$, on $\{\xi \in (a,b) \
, \ a < 0 \ , b > 0\}$. The joint density $f_{C,H,L}$ is simply $2
f_{c,h,l}$, restricted to  $\{\xi \in (0,b) \ , \ a < 0 \ , b >
0\}$. The two terms in the $j=0$ and second term in the $j=1$
summands vanish because they are independent of at least one of $a$
and $b$.

To calculate $E[C H L^2]$, the contribution of each summand in
(\ref{Siegmformula}) will be integrated in three univariate steps.
The first step will integrate over $a \in (-\infty,0)$ the product
of $a^2$ and the pertinent mixed second derivative. ${\partial \over
{\partial a}} \phi(\xi+Ka+Mb) da$ is to be interpreted as the
integration-by-parts element $d \phi(\xi+Ka+Mb)$, viewed as a
function of $a$.
\begin{eqnarray}
& & \int_{-\infty}^0 { \partial \over { \partial b}} a^2 {\partial
\over {\partial a}} \phi(\xi+Ka+Mb) da
\nonumber \\
& = & {2 \over K^2} {\partial \over {\partial b}}[\phi(\xi+M b)+(\xi+M b)\Phi(\xi+ M b)]
\mbox{ (for } K>0) \nonumber \\
& = &{{2 M} \over K^2} \Phi(\xi+ M b) \hspace{1.2cm} \mbox{ (for } K>0) \nonumber \\
& = &{{2 M} \over K^2} \Phi(\xi+ M b)-{{2 M} \over K^2} \mbox{ (for } K<0)
 \label{Siegmstep1}
\end{eqnarray}

Now expression (\ref{Siegmstep1}) will be multiplied by $\xi$ and
integrated over $\xi \in (0,b)$. For $K>0$ ($K<0$) it is convenient to
integrate $\Phi^*$ ($\Phi$).
These terms appear in (\ref{Siegmstep2}) and (\ref{Siegmstep3}).
The free term in (\ref{Siegmstep1}) contributes ${{2 M} \over K^2}
{b^2 \over 2}$ and cancels with the corresponding $b^2$ term in (\ref{Siegmstep3}).
\begin{eqnarray}
& & \int_0^b \xi {\partial \over {\partial b}} \int_{-\infty}^0 a^2
\phi(\xi + K da + M b) d\xi \nonumber \\
&=& {{2 M} \over K^2} \int_{M b}^{(M+1) b} y \Phi(y) dy -{{2 M^2 b}
\over K^2} \int_{M b}^{(M+1) b} \Phi(y) dy \nonumber \\
&=& {M \over K^2}[(M^2 b^2+1)\Phi(M b)-((M^2-1) b^2+1) \Phi((M+1) b) \nonumber \\
& &
+ M b \phi(M b)- (M-1) b \phi((M+1) b)] \nonumber \\
&=& -{M \over K^2}[(M^2 b^2+1)\Phi^*(M b)-((M^2-1) b^2+1) \Phi^*((M+1) b) \label{Siegmstep2} \\
& &
+M b \phi(M b)- (M-1) b \phi((M+1) b)]+{M \over K^2}b^2
\label{Siegmstep3}
\end{eqnarray}

Finally, expressions (\ref{Siegmstep2}) and (\ref{Siegmstep3}), multiplied by $b$ and
integrated over $b \in (0,\infty)$, via
\begin{equation} \label{Siegmstep4}
\int_0^\infty b^3 \Phi^*(A b) db = {3 \over {8 A^4}}
\ ; \ \int_0^\infty  b \Phi^*(A b) db = {1 \over {4 A^2}}
\ ; \ \int_0^\infty b^2 \phi(A b) db = {1 \over {2 A^3}}
\label{Siegmstep4}
\end{equation}
yield a rational function of $j$ (with $M=2 j$ and $K=-2 j$ or $K=-2(j-1)$) whose sum
contains only terms of the form $-\sum_1^\infty (-1)^j {1 \over j}=\log(2)$ and
$\sum_1^\infty {1 \over j^3}=\zeta(3)$, as in the statement of Lemma \ref{HL2C}.
Further details are omitted.

\section{Acknowledgements}

The topic under study was motivated by a project at ISTRA Research,
Israel. The collaboration of Shlomo Ahal, Jonathan Lewin and Alon
Wasserman is greatly appreciated. Ahal's careful reading and
constructive comments are an essential part of the paper. Warm
thanks are extended to my hosts at Columbia University's Statistics
Department during a sabbatical visit in the Spring of $2008$.

\begin{center}
Isaac Meilijson \\
{\em School of Mathematical Sciences} \\
{\em Raymond and Beverly Sackler Faculty of Exact Sciences} \\
{\em Tel-Aviv University , 69978 Tel-Aviv, Israel} \\
{\em E-mail: \tt{MEILIJSON@MATH.TAU.AC.IL}} \\
\end{center}

\end{document}